\documentclass[slac_one]{revtex4}
\usepackage{graphicx}
\usepackage{fancyhdr}
\begin{document}

\title{The SuperNEMO Experiment}

\author{R. B. Pahlka (for the SuperNEMO collaboration)}
\affiliation{The University of Texas at Austin, Austin, TX 78712, USA}

\begin{abstract}

The observation of neutrino oscillations has proven that neutrinos have mass.  This is direct evidence of physics beyond the Standard Model.  This discovery has renewed interest in neutrinoless double beta decay ($0\nu\beta\beta$) experiments which provide the only practical way to determine whether neutrinos are Majorana or Dirac particles.  Such experiments also have the potential to determine the absolute scale of the neutrino mass and help resolve the neutrino mass hierarchy question.  The NEMO-3 (Neutrino Ettore Majorana Observatory) is currently one of the most sensitive searches for neutrinoless double beta decay.  The main goal of SuperNEMO is to extend the sensitivity of the NEMO-3 search for neutrinoless double beta decay (and to measure two-neutrino double beta $2\nu\beta\beta$ decay).  The two isotopes under consideration for SuperNEMO are $^{82}$Se and $^{150}$Nd.  The target sensitivity is a $0\nu\beta\beta$ decay half-life at the level of $10^{26}$ years which will explore the degenerate neutrino mass hierarchy down to 50 meV.

\end{abstract}

\maketitle

\thispagestyle{fancy}

\section{INTRODUCTION}

Double beta decay is a rare nuclear process in which two neutrons in the same nucleus are spontaneously converted into two protons.  This process occurs when normal beta decay is either forbidden or suppressed, and may happen in at least two different ways:  by the emission of two electrons and two anti-electron neutrinos (already observed in many nuclei), or possibly by the emission of two electrons, but no neutrinos.

The recent observation of neutrino oscillations \cite{Mohapatra} and the resulting measurements of the neutrino mass differences has motivated experimental searches for the absolute neutrino mass \cite{Mohapatra}.  Neutrinoless double beta decay ($0\nu\beta\beta$) is the only practical way to understand the nature of neutrino mass and one of the most sensitive probes of its absolute value. If discovered, $0\nu\beta\beta$ would imply that the neutrino is a Majorana particle.  Ettore Majorana proposed that neutrinos could be their own anti-particles \cite{Majorana1937}, and this lead to Furry's conclusion \cite{Furry1939} that neutrinoless double beta decay is possible via neutrino exchange if the neutrinos are Majorana particles and have non-zero mass.

The effective $0\nu\beta\beta$  decay half-life $T_{1/2}^{0\nu}$ is proportional to the square root of the Majorana neutrino mass $\langle m_{\beta\beta}\rangle$ in equation (\ref{equ:2b0n}), where $G^{0\nu}$ is the kinematic phase-space factor and $M_{0\nu}$ is the nuclear matrix element. The experimental signature of $0\nu\beta\beta$ is two electrons originating in the same location, with the energy sum equaling the $Q_{\beta\beta}$ of the decay with a daughter nucleus in a final state. There are other mechanisms to explain neutrinoless double beta decay \cite{Mohapatra}, but the exchange of massive neutrinos (above) is the most favored due to the minimal required modifications to the Standard Model. 

\begin{equation}
[T_{1/2}^{0\nu}]^{-1} = G^{0\nu} \vert M_{0\nu} \vert ^{2} \langle m_{\beta\beta}\rangle^{2} \label{equ:2b0n} \end{equation}

There are several consequences for physics if the neutrino is found to be a Majorana particle.  It would imply lepton conservation violation, it would provide a natural explanation of the small mass of the neutrino and it could help explain the matter/anti-matter asymmetry in the Universe through leptogenesis \cite{Mohapatra}.


\section{THE SUPERNEMO DETECTOR}

SuperNEMO is a $\sim$100~kg source isotope ($^{82}$Se or $^{150}$Nd), tracker + calorimeter detector with a target neutrinoless double beta decay half-life sensitivity of $10^{26}$ years. The project is currently in a 3 year design study and R\&D phase and the collaboration comprises over 90 physicists from 9 countries. The R\&D program focuses on four main areas of study: isotope enrichment, tracking detector, calorimeter, and ultra-low background materials production and measurements.

\begin{figure}[htp]\centering
\includegraphics[height=10pc]{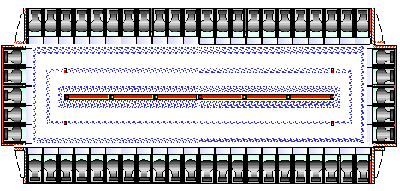}\hspace{3pc}
\includegraphics[height=10pc]{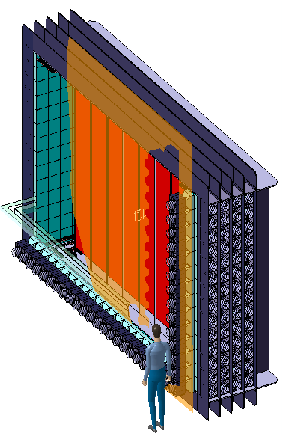}
\caption{ Preliminary design of the SuperNEMO detector: view from above (left) and three-dimensional cut-away view (right). } \label{fig:resolution} \end{figure}

\section{SuperNEMO detector design}

To fulfill these physics goals, SuperNEMO will build upon the NEMO-3 technology choice of combining calorimetry and tracking.  This gives the ability to measure individual electron tracks, vertices, energies and time of flight, and to reconstruct fully the kinematics and topology of an event.  Particle identification of gamma and alpha particles, as well as distinguishing electrons from positrons with the help of a magnetic field, form the basis of background rejection.  An important feature of NEMO-3 which is kept in SuperNEMO is the fact that the double beta decay source is separate from the detector, allowing several different isotopes to be studied. 

SuperNEMO will consist of about twenty identical modules, each housing around 5--7~kg of isotope.  A preliminary design of a SuperNEMO detector module is shown in Fig.~\ref{fig:resolution}.  The source is a thin ($\sim 40~mg/cm^2$) foil inside the detector.  It is surrounded by a gas tracking chamber followed by calorimeter walls.  The tracking volume contains more than 2000 wire drift chambers operated in Geiger mode, which are arranged in nine layers parallel to the foil.  The calorimeter is divided into $\sim$1000 blocks which cover most of the detector outer area and are read out by low background photomultiplier tubes (PMT).

\section{Status of the SuperNEMO R\&D Program}
The SuperNEMO collaboration was formed in 2005 with the goal of carrying out a three-year design study program (2006--2009) and producing a Technical Design Report (TDR) as outcome.  The SuperNEMO R\&D program is currently underway, supported by funding agencies in 9 countries. 

The expected improvement in performance of SuperNEMO compared to its predecessor NEMO-3 is shown in Table~\ref{tab:nemo3_comparison} which compares the parameters of the two experiments.  The most important design study tasks are described in the sections that follow. 

\begin{table}[tb]
\caption{ Comparison of the main NEMO-3 and SuperNEMO parameters. \label{tab:nemo3_comparison}}
{\begin{tabular}{@{}lcc@{}}\toprule

& NEMO 3 & SuperNEMO  \\ \colrule

isotope & $^{100}$Mo & $^{150}$Nd or $^{82}$Se \\  
mass    & 7 kg & 100--200 kg \\
signal efficiency & $18\%$ & $>30\%$ \\
$^{208}$Tl in foil & $<20 \mu$Bq/kg  & $<2 \mu$Bq/kg \\
$^{214}$Bi in foil & $<300 \mu$Bq/kg  & $<10 \mu$Bq/kg ($^{82}$Se) \\
energy resolution at 3 MeV & $8\%$ (FWHM)  & $4\%$ (FWHM) \\
half-life  & $T_{1/2}^{0\nu}>2\cdot 10^{24}$ years & $T_{1/2}^{0\nu}>(1-2)\cdot 10^{26}$ years \\
neutrino mass & $\langle m_{\beta\beta} \rangle < 0.3 - 0.7$ eV & $\langle m_{\beta\beta} \rangle < 40 - 110$ meV \cite{Caurier}\cite{Rodin} \\ \botrule
\end{tabular}}
\end{table}

\subsection{Calorimeter R\&D}

SuperNEMO aims to improve the calorimeter energy resolution to $7\%/\sqrt{E}$ FWHM at 1 MeV (4~\% @ the $Q_{\beta\beta}$ energy).  To reach this goal, several ongoing studies are investigating the choice of calorimeter parameters  such as scintillator material (organic plastic or liquid), and the shape, size and coating of calorimeter blocks.  These are combined with dedicated development of PMTs with low radioactivity and high quantum efficiency.  

Exceptional resolutions of 6.5\% at 1 MeV were measured for small PVT scintillators coupled to high QE PMTs. The SuperNEMO baseline design calls for large scintillator blocks ($~20 \times 20$~cm). Scintillators of this size read out through a lightguide showed an energy resolution of 9--10\% at 1 MeV.  Better results have been achieved by casting a large plastic hexagonal shaped scintillator directly on a hemispherical 8" PMT. With this configuration we have been able to reach the important milestone of 7--8\% $1/\sqrt{E}$ MeV energy resolution for the baseline detector design. Consequently the R\&D on solid scintillators will be focusing on cast scintillator solutions rather than lightguides to increase the light collection efficiency. The development program will also move away from the previous square-block designs and focus on more realistic hexagonal scintillator geometries. We note that there is room for further improvements by using a higher QE PMTs and more efficient scintillators \cite{Kauer}.

Liquid scintillator provides an alternative while maintaining good resolution (7--8\% at 1 MeV) and improving gamma tagging efficiency, but achieving the required resolution with large blocks as well as the engineering of the mechanical design and safety remain a challenge. The hybrid solution creates a more robust containment setup for the liquid, but achieving $<$7\% is very challenging. 

Long scintillator bars design can potentially give a more efficient detector with more background rejection power. It will drastically reduce the number of PMTs and facilitate a more compact detector design. Measurements so far yield 10\% at 1 MeV. Work is in progress to understand if this resolution can be improved to 8--9\% and whether a worse energy resolution can be compensated by the above advantages of this detector configuration. 

Last years have seen a significant progress in development of novel photo-detectors. PMTs with a QE of over 40\% are now available. Using the latest achievements in PMT, reflector, and scintillator technology the SuperNEMO collaboration has demonstrated the feasibility of achieving the target energy resolution necessary to reach the sensitivity goal of the experiment. The remaining challenge is to demonstrate that the achieved energy resolution can be maintained at the mass production scale. SuperNEMO expects to make the final decision on the calorimeter design in mid-2009. The large scale construction will start in 2011 \cite{Kauer}.

\subsection{Tracker design}

The SuperNEMO tracker consists of octagonal wire drift cells operated in Geiger mode.  Each cell is around 4~m long and has a central anode wire surrounded by 8--12 ground wires, with cathode pickup rings at both ends.  Signals can be read out from the anode and/or cathodes to determine the position at which the ionising particle crossed the cell.  

The tracking detector design study looks at optimising its parameters to obtain high efficiency and resolution in measuring the trajectories of double beta decay electrons, as well as of alpha particles for the purpose of background rejection.  The tracking chamber geometry is being investigated with the help of detector simulations to compare the different possible layouts.  In addition, several small prototypes have been built to study the drift chamber cell design and size, wire length, diameter and material, and gas mixture \cite{Nasteva}. 

The first 9-cell prototype was successfully operated with three different wire layouts, demonstrating a plasma propagation efficiency close to 100\% over a wide range of voltages \cite{Nasteva}.  In addition, a 90-cell prototype has recently been constructed which will test the mechanics and large-scale operation of the drift cell system.  A SuperNEMO module will contain several thousand drift cells with 8--12 wires each.  The large total number of wires requires an automated wiring procedure, thus a dedicated wiring robot is being developed for the mass production of drift cells.  

\subsection{Choice of source isotope}
The choice of isotope for SuperNEMO is aimed at maximising the neutrinoless signal over the background of two-neutrino double beta decay and other nuclear decays mimicking the process.  Therefore the isotope must have a long two-neutrino half-life, and high endpoint energy and phase space factor $G_{0\nu}$ ($T_{1/2}^{0\nu} \sim G_{0\nu}^{-1}$).  The enrichment possibility on a large scale is also a factor in selecting the isotope.  The main candidate isotopes for SuperNEMO have emerged to be $^{82}$Se and $^{150}$Nd.  A sample of 4~kg of $^{82}$Se has been enriched and is currently undergoing purification.  The SuperNEMO collaboration is investigating the possibility of enriching large amounts of $^{150}$Nd via the method of atomic vapor laser isotope separation.  However, the collaboration has not ruled out other possible sources. 

\subsection{Radiopurity of the source}
SuperNEMO will search for a very rare process, therefore it must maintain ultra-low background levels.  The source foils must be radiopure, and their contamination with naturally radioactive elements must be precisely measured. The most important source contaminants are $^{208}$Tl and $^{214}$Bi, whose decay energies are close to the neutrinoless signal region.  SuperNEMO requires source foil contamination to be less than $2 \mu Bq/kg$ for $^{208}$Tl and less than $10 \mu Bq/kg$ for $^{214}$Bi
In order to evaluate these activities, a dedicated BiPo detector was developed which can measure the signature of an electron followed by a delayed alpha particle.  The first BiPo prototype (BiPo1) was installed in the Modane Underground Laboratory in February 2008 and is currently running with 20 modules.  The objective for this prototype is to measure the backgrounds and surface contamination of the prototype's plastic scintillators.  After three months, an upper limit on the sensitivity for a standard $12 m^2$ BiPo detector was calculated to be less than $7.5 \mu Bq/kg$ for $^{208}$Tl (90\% C. L.) \cite{Bongrand}. 

A second BiPo prototype (BiPo2) is also in final stages of characterization.  This prototype uses large, thin (75 cm x 75 cm x 1 cm) sheets of Bicron plastic scintillator whereby a source foil can be inserted between two sheets.  The signal is read out on the sides with arrays of PMTs coupled to custom light guides.  The BiPo2 prototype was installed in the Modan Underground Laboratory in July 2008.  Measurements are currently underway for surface and bulk scintillator contamination and optical cross-talk.    

\section{Conclusion}
An extensive R\&D program is underway to design the next-generation neutrinoless double beta decay experiment SuperNEMO. It will extrapolate the successful technique of calorimetry plus tracking of NEMO-3 to 100--200~kg of source isotope, aiming to explore the degenerate neutrino mass hierarchy.  Due to its modular approach, SuperNEMO can start operation in stages, with the first module installed as early as 2011, and all twenty modules running by 2013--2014. 

\section{Acknowledgements}
The collaboration would like to acknowledge support by STFC (UK), the Grants Agency of the Czech Republic, RFBR (grant \# 06-02-16672, Russia), NSF (grant PHY-0702695, USA), and the ICHEP 2008 Conference for student registration fee support.

\end{document}